\documentclass[dvips]{arxstspdf}
\usepackage{flushend}

\volume{23}
\issue{3}
\pubyear{2008}
\firstpage{325}
\lastpage{331}
\doi{10.1214/08-STS244REJ}
\referstodoi{10.1214/07-STS244}

\begin{document}
\begin{frontmatter}

\title{Rejoinder: Quantifying the Fraction of Missing Information
for Hypothesis Testing in Statistical and Genetic Studies}
\runtitle{Rejoinder}

\begin{aug}
\author[a]{\fnms{Dan L.} \snm{Nicolae}\ead[label=e3]{nicolae@galton.uchicago.edu}},
\author[b]{\fnms{Xiao-Li} \snm{Meng}\corref{}\ead[label=e2]{meng@stat.harvard.edu}} \and
\author[c]{\fnms{Augustine} \snm{Kong}\ead[label=e1]{kong@decode.is}}
\runauthor{D. L. Nicolae, X.-L. Meng and A. Kong}

\affiliation{The University of Chicago, Harvard University and deCode Genetics}

\address[a]{Dan L. Nicolae is Associate Professor, Departments of
Medicine and Statistics,
The University of Chicago, Chicago, Illinois 60637, USA (\printead{e3}).}
\address[b]{Xiao-Li Meng is Whipple\break V. N. Jones Professor and Chair of Statistics,
Harvard University, Cambridge, Massachusetts, USA (\printead{e2}).}
\address[c]{Augustine Kong is Vic\`{e} President of Statistics, deCode Genetics,
Sturlugata 8, IS-101 Reykjavik, Iceland (\printead{e1}).}
\end{aug}

% ABSTRACT

% KEYWORDS
%
%genetic linkage studies, haplotype-based association studies, Kullback--Leibler
%information, noninformative prior, partial likelihood, relative information}
\end{frontmatter}

%s1 ###
\section{A Professional Joy}\label{sec:intro}

Few authors would not be pleased when discussants implement their
methods or follow-up on their ideas. It is therefore a professional
joy to see every discussant doing both! Our heartfelt thanks go to
all discussants, and to the Executive Editor, Ed George, for
bringing us such joy!

Incidentally, the three discussions cover nicely the three main
parts of our paper. Zheng and Lo's discussion centers on our
motivating application, namely, designing follow-up strategies in
genetic studies, but with the additional consideration of the
uncertainty in the measures themselves. Doss's discussion focuses on
the second part of our paper, namely, the likelihood-based relative
measure, but with applications to survival analysis where the use of
partial likelihood \mbox{reveals} very interesting (and inevitably
confusing) complications. Chang, Chen, Chien and Hsing (hereafter
$\mathrm{C}_{3}\mathrm{H}$) comment on the third part of our paper, the Bayesian measures
for small samples, and implement variations that are applied to
problems in infectious disease research and isotonic regression.

Our responses are organized in the aforementioned order. We very
much appreciate all the key messages conveyed by the discussants,
though for a few of them we offer alternative explanations. Some
questions posed by the discussants make nice Ph.D. or master thesis
topics, so we summarize them at the end of this rejoinder.

%s2 ###
\section{Zheng and Lo: Design with Uncertainty}

Zheng and Lo further emphasize the critical role of measuring
relative information in designing follow-up studies, and touch upon
the issue of optimal design under a given measure. In particular,
they consider a setting with multiple variables, and suggest an
extension of our harmonic rule (19) for combining multiple studies
to the setting of combining multiple variables. Since our rule (19)
was derived under the assumption that individual studies are
independent, we surmise that Zheng and Lo's setting is under similar
considerations, where variables are considered to be independent of
each other and their contributions to the overall log-likelihood are
additive. Otherwise we will need to consider all variables jointly
in measuring relative information. Nevertheless, it would be useful
to investigate how Zheng and Lo's combining rule (1) performs as a
quick approximation to the measure that uses the full likelihood,
when the independence assumption fails. Zheng and Lo's (1) could be
quite appealing to a practitioner who chooses to deal with multiple
variables separately, especially for testing purposes, because of
the technical difficulty in specifying a reliable large joint
multivariate model.

Zheng and Lo also correctly point out that the actual test
statistics (e.g., log-likelihood ratio) from a follow-up study can
be quite different from what is predicted by our measures of
relative information, $\mathcal{R}I_1$ and $\mathcal{R}I_0$. There
are several
different ways of investigating this uncertainty. Zheng and Lo take
a direct approach, by simulating the actual ratio of complete-data
log-likelihood ratio versus the observed-data log-likelihood ratio,
which they denote by $\mathcal{R}I_y$, as a function of the missing
data. The
simulations are done by drawing the missing data from the
conditional distribution given the observed data and the parameter
value estimated by the observed-data MLE. In the binomial example, a
simulation study is used to \vspace*{2pt} demonstrate that
$\mathcal{R}I_1^{-1}$
is the average of
$\mathcal{R}I_y^{-1}$,
which itself exhibits considerable variation.

Here we wish to point out a subtlety. Whereas
$\mathcal{R}I_1^{-1}$
has the nice interpretation of being the ratio of the \textit{expected}
complete-data lod score to the observed-data lod score, this
expectation is calculated under the assumption that the value of the
parameter under the alternative hypothesis is the same as the one
under which the (conditional) expectation is calculated. There is no
confusion about this assumption when the alternative hypothesis is
sharp, that is, when it has a fixed known value. This is essentially
what Zheng and Lo assumed, as they considered a number of
alternative values ($p=0.525, 0.55, 0.65$) for their simulation
studies. It is clear that under such a setting,
$E[\mathcal{R}I_y^{-1}|Y_{\mathrm{ob}}; \theta =\theta _{\mathrm{ob}}]=\mathcal{R}I_1^{-1}$,
by the definition of $\mathcal{R}I_1$.

However, once we move away from this setting and allow the use of
the actual complete-data lod score
${\operatorname{lod}}(\theta_{\mathrm{co}}, \theta _0|Y_{\mathrm{co}})$,
where
$\theta_{\mathrm{co}}$
is the complete-data MLE, then things can
become much more complicated. For example,
$E[\mathcal{R}I_y^{-1}|Y_{\mathrm{ob}}; \theta =\theta_{\mathrm{ob}}]=\mathcal{R}I_1^{-1}$
no longer holds because in general,
\begin{eqnarray}\label{eq:note}
\nonumber&&\mathrm{E}[{\operatorname{lod}}(\theta _{\mathrm{co}},\theta_0|Y_{\mathrm{co}})|Y_{\mathrm{ob}};
\theta=\theta_{\mathrm{ob}}]
\\[-8pt]\\[-8pt]
\nonumber&&\quad\not=\mathrm{E}[{\operatorname
{lod}}(\theta_{\mathrm{ob}},\theta_0|Y_{\mathrm{co}})|Y_{\mathrm{ob}}; \theta =\theta_{\mathrm{ob}}].
\end{eqnarray}
Mathematically, our key identity (13) requires both $\theta _1$ and
$\theta _2$ to be fixed known constants (given the observed data), so
one cannot take
$\theta _1=\theta_{\mathrm{co}}$,
which would be a random variable,
even after conditioning on
$Y_{\mathrm{ob}}$.
This technical requirement,
however, is a reflection of a more fundamental difficulty in
measuring (relative) information. If the additional data change the
MLE (i.e., from $\theta_{\mathrm{ob}}$ to $\theta_{\mathrm{co}}$), which can be viewed as a
``center'' of the likelihood, then measuring relative information,
in terms of relative strength against a null hypothesis, becomes a
very tricky task. Perhaps this is more clearly seen by viewing the
likelihood function as an un-normalized posterior density, and
imagining that there are two posterior densities. One is centered
around a value close to $\theta _0$ with a small posterior variance
(i.e., the one based on $Y_{\mathrm{co}}$) and the other is centered
around a
value farther away from $\theta_0$ but also with larger spread (i.e.,
the one based on $Y_{\mathrm{ob}}$). It is then debatable how to
compare the two
posteriors' respective strengths in discrediting the value of
$\theta_0$; certainly it is a much harder task than when both
posteriors are centered at the same location.

With our measures we circumvent this problem by first calculating
the log-likelihood ratio or lod score for the same null value
$\theta_0$ and same alternative value $\theta_1$, given both the observed
data and complete data. We then estimate the unknown value of
$\theta_1$, or even $\theta_0$ when the null is not sharp, by the MLE
under the alternative and null hypotheses, respectively.
Alternatively, as we demonstrated via the simple binomial example,
when the complete-data likelihood is from an exponential family
[which is the case for the binomial when $p$ is restricted to
(0, 1)], what we proposed was to measure how anti-conservative our
test would be if we imputed the complete-data sufficient statistics
under the alternative hypothesis and then pretended that they were
real data (for $\mathcal{R}I_1$), or how conservative our test procedure
would be if we imputed under the null and then pretended that they
were real data (for $\mathcal{R}I_0$).

In that sense, the only uncertainty in our measures is the
uncertainty caused by using the observed-data MLEs for $\theta _1$ and
$\theta _0$. This is different from Zheng and Lo's simulation and
variance calculation, which attempts to capture the conditional
variation in $\mathcal{R}I^{-1}_y$ given the observed data. However,
it is
important to point out that, because Zheng and Lo's setting treats
the alternative value of the hypothesis as known, their variation is
also different from the actual (conditional) variation in the ratio of the
complete-data lod score and the observed-data lod score. The latter
would be
%
%e1 ###
\begin{equation}
\frac{\operatorname{Var}[{\operatorname{lod}}
(\theta_{\mathrm{co}}, \theta_0|Y_{\mathrm{co}})|Y_{\mathrm{ob}},\theta] }{
{\operatorname{lod}}^2(\theta_{\mathrm{ob}}, \theta_0|Y_{\mathrm{ob}})},
\end{equation}
which then can be evaluated at $\theta =\theta_{\mathrm{ob}}$, as
Zheng and Lo
suggested. Which of these variance calculations is most relevant for
practical purposes is worthy of exploring, and we thank Zheng and Lo
for their recognition of this issue.

It is worth reiterating here that the range of genetics/genomics
applications of the proposed measures of information is expanding
with every high-throughput technology that is developed in this
rapidly moving field. For example, in many applications, the
individual genotypes on the genome are not measured
deterministically; instead, a distribution on all possible states is
inferred from the raw data. Examples of this include: (i) genotype
calling using data from the new sequencing technologies such as
those from Solexa and Applied Biosystems, where uncertainty in calls
comes from technical errors, sequence assembly and sequence
similarity (\cite*{brockman08}); (ii) imputation of genotypes for
untyped markers using information from a reference database such as
HapMap, where uncertainty is caused by imperfect prediction and by
the size of the training data set (\cite*{nicolae06}); and (iii)~calling
genotypes of Copy Number Variation (CNV), where the
variability is caused by uncertainty in the boundaries of the CNVs
and by technical variability in the probe measurements
(\cite*{redon06}). In all of these situations, instead of data
yielding a
genotype,~$G$, the raw information is processed into a distribution
on all possible values for $G$, $P(G|{\mathrm{data}})$. These
distributions can be used, for example, in testing for genetic
association of a disease or quantitative trait with the marker under
investigation. The measures proposed in our paper can be applied
directly (similarly to the haplotype application presented in the
paper) to quantify the amount of information relative to having
observed the genotypes. The measures are important because it is
possible, with additional laboratory work, to determine the
genotypes with certainty. The complications arise when information
on different markers that are in the same biological unit (such as
a gene or a pathway) are combined into a single association test. This
is the case where the discussion above is relevant and further
research is necessary.

%s3 ###
\section{Doss: So What Went Wrong with Partial Likelihood?}

We very much appreciate Doss's exploration of applying our measures
to the survival analysis setting, and were very intrigued by the
problems he reported with Cox's partial likelihood. As we stated in
the first section of our paper, one basic requirement in measuring
relative information is that we need to assume that the procedure
under investigation is ``optimal'' in some sense (e.g., being
full-likelihood based). This requirement is needed to prevent
paradoxical situations where less data can lead to more information,
much like the ``self-efficient'' requirement in \citet{Meng94}. A
good illustration of such a situation is a least-square regression
in which the variance depends on the value of the covariate. While
the ordinary least-square estimators enjoy the robustness in the
sense of still being consistent in the presence of
heteroscedasticity, they are not self-efficient (\cite*{Meng94})
because one can have a much more efficient least-square estimator
with fewer data if the additional data happen to be those with much
higher variances; see \citet{meng01} for a detailed illustration. So
Doss's finding, that $\mathcal{R}I_1$ may not be less than 1 for some of
the data sets he used, reminded us to look into the possibility that
the partial likelihood approach may fail this basic requirement.

When ``partial likelihood'' is taken to mean literally any part of a
full likelihood, this failure is obvious, because it would be
trivial to construct many examples where the part chosen is so
inefficient compared with the full likelihood that
``self-efficiency'' cannot possibly hold (even taking into account
that ``self-efficiency'' is a weaker requirement than the usual full
efficiency). So the question of real interest here is what happens
in the specific case of Cox's partial likelihood for the
proportional hazard model, an approach that is often considered to
produce results as good as the full likelihood method, at least for
practical purposes. The answer to this question, however, is not
straightforward.

The simplest situation is when there is no censoring, in which case
it is known that Cox's partial likelihood for the proportional
hazard model is also a genuine likelihood based on part of the data,
that is, on the ranks of all the observed failure times
(\cite*{FlemingHarrington91}, Chapter 4). Since it is a genuine likelihood, it must
be self-efficient, and there should be no problem to apply our (16)
or any subsequent formulas, as long as they are implemented
correctly (see below). When there is censoring, the discussion in
\citet{FlemingHarrington91} shows that a further sacrifice of
efficiency is needed in order to arrive at Cox's partial likelihood
via the rank-data formulation. Currently we are unable to determine
the impact of this further sacrifice on self-efficiency.

What we are able to determine, or rather to detect, however, is that
there is another reason that can explain Doss's ``surprising
findings,'' even if the self-efficiency issue is not relevant. The
problem lies in how one defines \textit{observed data}, and by
comparison, what constitutes \textit{complete data}. One might find
this is a rather odd inquiry---how hard could it be to determine
what is observed and what is missing?

To see why this can be a problem, let us set up the notation
carefully. Using Doss's ${\mathrm D}$ notation for data, we distinguish
three data sets: $\mathrm{D}_{\mathrm{full}}$ is the full data set
that would be observed
if there were no censoring, $\mathrm{D}_{\mathrm{cens}}$ is the
available/observed
censored data,
and $\mathrm{D}_{\mathrm{part}}$ is Cox's partial data, that is, the
actual data
used for calculating Cox's partial likelihood function.

Given this setup, we can use $\mathcal{R}I_1$ to measure the loss of
information due to censoring by setting
$\{Y_{\mathrm{ob}}=\mathrm{D}_{\mathrm{cens}},Y_{\mathrm{co}}=\mathrm{D}_{\mathrm{full}}\}$,
using our generic
notation; we believe Doss's first
reported $\mathcal{R}I_1$ value, $0.987$, is for this purpose. We can also
measure the loss of information from using the partial likelihood
approach compared with the full-likelihood approach, which
corresponds to setting $\{Y_{\mathrm{ob}}=\mathrm{D}_{\mathrm
{part}}, Y_{\mathrm{co}}=\mathrm{D}_{\mathrm{cens}}\}$. Doss does not
seem to provide such a measure. We remark that we may also measure
the loss of information of using $Y_{\mathrm{ob}}=\mathrm{D}_{\mathrm{part}}$ compared with using
$Y_{\mathrm{co}}=\mathrm{D}_{\mathrm{full}}$, though this $\mathcal{R}I_1$ may not be numerically the same
as the product of the previous two because they assume different
observed data in computing the MLEs and take different conditional
expectations over the missing data.

The setting Doss provided is, however, more complicated. Imagine
that we had collected additional samples, possibly censored. Let
$\mathrm{D}_{\mathrm{cens}}^{aug}$ denote the augmented data set that
includes $\mathrm{D}_{\mathrm{cens}}$;
$\mathrm{D}_{\mathrm{cens}}\subset\mathrm{D}_{\mathrm{cens}}^{aug}$.
We then obviously can ask what is the relative information in
$Y_{\mathrm{ob}}=\mathrm{D}_{\mathrm{cens}}$ compared with the
augmented sample $Y_{\mathrm{co}}=\mathrm{D}_{\mathrm{cens}}^{aug}$. This
is, we believe, what Doss intended. However, since Cox's partial
likelihood is a very popular approach, Doss wanted to measure the
relative information when using the partial likelihood, not the full
likelihood.

Because Cox's partial likelihood uses the partial data $\mathrm
{D}_{\mathrm{part}}$, we
then should set $\{Y_{\mathrm{ob}}=\mathrm{D}_{\mathrm{part}},
Y_{\mathrm{co}}={\mathrm D}_{\mathrm{part}}^{aug}\}$, where ${\mathrm
D}_{\mathrm{part}}^{aug}$ is
Cox's partial data from the augmented sample $\mathrm{D}_{\mathrm
{cens}}^{aug}$. That is,
the moment we decide to measure the relative information for using
Cox's partial likelihood approach, our relative information is
\textit{no longer} about $Y_{\mathrm{ob}}=\mathrm{D}_{\mathrm{cens}}$
relative to $Y_{\mathrm{co}}=\mathrm{D}_{\mathrm{cens}}^{aug}$, but
rather about $Y_{\mathrm{ob}}=\mathrm{D}_{\mathrm{part}}$ relative
to $Y_{\mathrm{co}}={\mathrm D}_{\mathrm{part}}^{aug}$, because the
latter are the actual data sets used by the Cox regression.

Recognizing the correct $Y_{\mathrm{ob}}$ and $Y_{\mathrm{co}}$
directly affects how we
compute, among other things, the denominator of $\mathcal{R}I_1$. With
$Y_{\mathrm{ob}}=\mathrm{D}_{\mathrm{part}}$ and $Y_{\mathrm{co}}={\mathrm D}_{\mathrm{part}}^{aug}$, the conditional expectation called
for by the denominator of $\mathcal{R}I_1$ of (18) in our paper should be
with respect to
%
%e2 ###
\begin{equation}\label{eq:corr}
f(Y_{\mathrm{co}}|Y_{\mathrm{ob}}; \theta_{\mathrm{ob}})=f({\mathrm
D}_{\mathrm{part}}^{aug}|\mathrm{D}_{\mathrm{part}};\theta_{\mathrm{ob}}).
\end{equation}
However, the conditional distribution Doss actually used in his
Monte Carlo simulation appears to be
%
%e3 ###
\begin{equation}\label{eq:twod}
f(\tilde{Y}_{\mathrm{co}}|\tilde{Y}_{\mathrm{ob}};
\theta_{\mathrm{ob}})=f(\mathrm{D}_{\mathrm{cens}}^{aug}|\mathrm
{D}_{\mathrm{cens}};\theta_{\mathrm{ob}}).
\end{equation}
The critical difference between~(\ref{eq:corr}) and~(\ref{eq:twod})
is in what is being conditioned upon, namely, $\mathrm{D}_{\mathrm
{part}}$ versus
$\mathrm{D}_{\mathrm{cens}}$. (The difference between $Y_{\mathrm{co}}$ and $\tilde Y_{\mathrm{co}}$ is less
important here because ${\mathrm D}_{\mathrm{part}}^{aug}$ is a
deterministic function of
$\mathrm{D}_{\mathrm{cens}}^{aug}$, so if we can calculate or
simulate with respect to a
correctly specified conditional distribution of $\mathrm{D}_{\mathrm
{cens}}^{aug}$, then we
can do so for any of its functions/margins.) We point
out this difference because the use of~(\ref{eq:corr}) is consistent
with our original definition, as it uses the same \textit{observed
data} set for both the numerator and denominator of $\mathcal{R}I_1$.
Using~(\ref{eq:twod}), however, will result in unclear consequences. For
one thing, our key inequality (16) is no longer guaranteed to hold
because the ``Kullback--Leibler information'' part would then be of
the form $\int p_1(x)\* \log[p_2(x)/p_0(x)]\mu(dx)$, which is not
guaranteed to be nonnegative when $p_1(x)\not=p_2(x)$.

Doss's explanation of his ``surprising findings'' is also based on
an inconsistency, but it is the inconsistency between including some censored
observations for the denominator versus only using the uncensored
cases for the numerator. Our investigation above, however, reveals that the
problem lies in using the ranks of the failure times, as
in $\mathrm{D}_{\mathrm{part}}$ and ${\mathrm D}_{\mathrm
{part}}^{aug}$, which is not the same as using the failure times
themselves, as in $\mathrm{D}_{\mathrm{cens}}$ and $\mathrm
{D}_{\mathrm{cens}}^{aug}$. This difference is
irrespective of censoring, because even without censoring, in which
case $\mathrm{D}_{\mathrm{cens}}=\mathrm{D}_{\mathrm{full}}$, the
critical difference between the
conditioning in~(\ref{eq:corr}) and in~(\ref{eq:twod}) remains.

Intriguingly, the need for setting up notation carefully is
demonstrated by another more subtle difference between~(\ref{eq:corr})
and~(\ref{eq:twod}), at least when there is no
censoring. In both~(\ref{eq:corr}) and~(\ref{eq:twod}), we used the
generic notation $\theta_{\mathrm{ob}}$ to denote an estimator of
$\theta$ based
on the observed data. However, in the current setting, $\theta$
consists of both the parameter of interest, $\beta$, and the
(infinite-dimensional) nuisance parameter $\Lambda_0$, the baseline
cumulative hazard. This recognition immediately reveals a
problem for~(\ref{eq:corr}), because there is little information in
$\mathrm{D}_{\mathrm{part}}$ for estimating $\Lambda_0$. After all,
the most celebrated
feature of Cox's partial likelihood is its ability to estimate
$\beta$ without having to deal with $\Lambda_0$.

When there is no censoring, this problem also turns out to be the
solution because $f({\mathrm D}_{\mathrm{part}}^{aug}|\mathrm
{D}_{\mathrm{part}};\theta)$ is actually free of
$\Lambda_0$, a consequence of the fact that Cox's partial likelihood
is identical to the full likelihood of~$\beta$ based on the ranks
alone. One therefore can carry out~(\ref{eq:corr}) by calculating or
simulating with respect to $f({\mathrm D}_{\mathrm
{part}}^{aug}|\mathrm{D}_{\mathrm{part}};\beta=\beta_{\mathrm{ob}})$, where
$\beta_{\mathrm{ob}}$ is the Cox regression estimator based on
$\mathrm{D}_{\mathrm{part}}$.

When there is censoring, the picture becomes less clear,
because it is then possible for $f({\mathrm D}_{\mathrm
{part}}^{aug}|\mathrm{D}_{\mathrm{part}};\theta)$ to depend
on the baseline $\Lambda_0$. This is not a contradiction to the
celebrated feature of Cox's partial likelihood, that is, its robustness
to the
specification of $\Lambda_0$. The relative information $\mathcal{R}I_1$
itself may well depend on the actual distribution of the failure
time when there is censoring, because the probability of censoring
generally depends on the actual distribution of the failure time.
What this means is that whereas we can still define $\mathcal{R}I_1$
theoretically as we did, it cannot be estimated using $\mathrm
{D}_{\mathrm{part}}$
alone. This dilemma could be taken as a defense for using~(\ref{eq:twod}),
at least for practical purposes, especially
considering the difficulties in implementing~(\ref{eq:corr}) even if
$\theta$ is known.

However, to avoid the type of ``surprising findings'' that Doss found,
we would resolve this dilemma by nonetheless using~(\ref{eq:corr}) but
with the nuisance parameter $\Lambda_0$ estimated from $\mathrm
{D}_{\mathrm{cens}}$,
for instance using the Nelson--Aalen estimator used by Doss. That is,
$\mathrm{D}_{\mathrm{cens}}$
enters the calculation only through the estimation of $\Lambda_0$.
This dependence on $\mathrm{D}_{\mathrm{cens}}$ will not cause the
type of problems that
Doss reported, because it does not alter the conditioning as called for
by~(\ref{eq:corr}) and because our (18) permits its numerator and
denominator to depend on different parts of the same $\theta_{\mathrm{ob}}$. Of
course, this dependence makes uncertainty quantifications, such as
those emphasized by Zheng and Lo, even more important, as well
as more complicated, because $\Lambda_0$ is an infinite-dimensional nuisance parameter.

In a nutshell, all these complications remind us of the great caution
we must exercise once we deviate from the full-likelihood setting.
Indeed, whereas we recognized early the existence of an
alternative explanation of Doss's finding, one of our initial
explanations itself was a product of our lacking full appreciation of
the theoretical intricacy of Cox's partial likelihood. We are certainly
grateful to Doss for providing such a rich and intricate example,
even though, or perhaps especially because, we were nearly tripped
up by it!

We also very much appreciate Doss's attempt to generalize our measure
to the nonlikelihood setting. Indeed, our motivating examples, both
the toy example with the binomial distribution and the real genetic
applications, are for nonlikelihood types of testing, either with a
Wald-type test in the binomial case or with non-parametric lod
scores in the genetic setting. However, precisely for the
``non-self-efficient''
reason discussed above, it soon became clear to us
that in order to avoid paradoxical situations where fewer data may
lead to more information, we need to associate a test with a model
in order to proceed, as we did in Section 2.3.

If we understand Doss's notation correctly, his $\mathcal{R}I_w$ can be
obtained from our $\mathcal{R}I_1$ by first associating his tests with
normal models, and hence the likelihood ratio test is the same as
the Wald test. It is easy to verify that once we associate the
complete-data test with the normal model (i.e., pretending the
large-sample approximation is exact), the denominator of $\mathcal{R}I_1$ is
the same as the denominator of Doss's $\mathcal{R}I_w$ as given in
his (5). If
we further associate the observed-data test with the normal model,
then the numerators of $\mathcal{R}I_1$ and $\mathcal{R}I_w$ will
be the same, and
hence $\mathcal{R}I_w$ will be identical to $\mathcal{R}I_1$.

An astute reader might question why we need to associate the
normal model with the complete-data test and observed-data test
separately. Should not the complete-data model automatically imply
the observed-data model? The answer is ``yes'' if both the complete-data
test and the observed-data test are derived from a coherent
probability model (e.g., if both are likelihood ratio tests).
However, when tests are derived nonparametrically, or even
parametrically but without following the full-likelihood recipe
(for instance, using a partial likelihood), there is no guarantee that
the two
tests are ``coherent'' with each other in the sense that by
integrating out the missing values in the complete-data associated
model one would automatically obtain the observed-data associated
model. Indeed, Doss's $\mathcal{R}I_w$ can also exceed 1 if the
variance of the complete-data test statistic is larger than that of
the observed-data test statistic, a phenomenon that can occur with
an ordinary least square estimator, as discussed above. A logical
conclusion is then that even when $\mathcal{R}I_w$ seems to be ``likelihood
free,'' fundamentally its rationality is guaranteed only when a
(normal) likelihood family can be associated with it.

%s4 ###
\section{$\mathrm{C}_{3}\mathrm{H}$: Infectious Disease Studies and
Isotonic Regression}

We are pleased to see that $\mathrm{C}_{3}\mathrm{H}$ took on the task of implementing
our suggested Bayesian measures in the context of infectious disease
and regression. For infectious disease, $\mathrm{C}_{3}\mathrm{H}$'s goal was to decide
whether to invest in finding out the infectious times for the
existing cases for which only the removal times are known, or in finding
additional families/individuals whose removal times are known (but
whose infectious times are unknown). This consideration is important here
because identifying the infection time is typically much harder (if
possible at all) than identifying the removal time (e.g., death
time). For the isotonic regression application, $\mathrm{C}_{3}\mathrm{H}$
considered the
design issue: whether to add more measurements at the existing
design points or to add new design points that interlace with the
existing design points.

While we are excited by these new applications, we are somewhat
puzzled, and worried, by $\mathrm{C}_{3}\mathrm{H}$'s findings in both examples.
For the
infectious disease example, our intuition would suggest that
identifying \mbox{infection} times would be more important for testing
efficacy of vaccine than finding more individuals with only removal
times known, especially when it is not clear (at least to us from the
model description given by $\mathrm{C}_{3}\mathrm{H}$) whether ``removal'' here means
death or cure (and thus possible immunity). $\mathrm{C}_{3}\mathrm{H}$ gave an example
where the measured relative information in 20 households with
only removal times is about 80\% compared with the situation in which
everyone's infection time is also known. But it is only about 30\%
relative information compared with having four additional households
with removal times only. This sharp difference is a surprise to us,
and makes us wonder whether it is a reflection of issues with $\mathrm{C}_{3}\mathrm{H}$'s (BI3) or a
defect in implementation (e.g., failure of an MC algorithm).

Similarly, we are surprised to see that, in the context of testing
for monotonicity of a regression function, doubling the measurements
at existing design points creates substantially more information
than adding an equal amount of new design points interlaced with
existing design points. $\mathrm{C}_{3}\mathrm{H}$ gave an example where the
observed data
only have about 15\%
information relative to the former design, compared with 35\%
information relative to the latter design. This is rather
counterintuitive, because for
estimating a response surface with a fixed number of measurements,
it is often wise to spread out more
design points rather than to take more measurements on fewer design
points. For example,
for the simple linear regression $y_i=\beta x_i+\varepsilon_i$ (the one
that generated $\mathrm{C}_{3}\mathrm{H}$'s data), the variance of the least-square
estimator would be inversely proportional to $S_x=\sum_i x_i^2$; for
$\mathrm{C}_{3}\mathrm{H}$'s setting, $S_x=\sum_{i=0}^9 (i/9)^2= 95/27$.
Doubling the
number of measurements at each existing design point clearly will
double $S_x$:
$S_x=190/27=7.037$. On the other hand, $\mathrm{C}_{3}\mathrm{H}$'s second
design, if we
understand their description correctly, is to use $i/12,$ $i=1,\ldots,
5, 7, \ldots, 11$, as the additional 10 design points. Under this
design, $S_x=\sum_{i=0}^9 (i/9)^2 + \sum_{i=1}^{11} (i/12)^2 -
(6/12)^2=1465/216=6.78$. So while the first design is indeed
slightly better, the relative variance ratio is $96\%$, nowhere near
the 2.5-fold increase in information suggested by $\mathrm{C}_{3}\mathrm{H}$'s results
($0.346/0.139=2.5$).
Of course, we understand that $\mathrm{C}_{3}\mathrm{H}$ are measuring
information in testing, not estimation, and their method is far more
sophisticated than the simple linear regression. Nevertheless, we
find the 2.5-fold increase rather counterintuitive, and would be
very interested in seeing it confirmed independently in a different
way.

$\mathrm{C}_{3}\mathrm{H}$ also touch on the intricate issue of dealing with nuisance
parameters under the null. They
suggest two ways of averaging: either averaging the numerator and
denominator separately and then taking the ratio (BI3), or directly
averaging the ratio (BI4). Here all averaging is performed with respect to
the posterior distribution of the nuisance parameter under the null.
As we discussed in Section 6.3 (and elsewhere) of our paper, dealing
with nuisance parameters is a complicated issue, even with the
Bayesian approach, because we do not have reliable priors for them,
nor do we know enough about the sensitivity of these measures,
including $\mathrm{C}_{3}\mathrm{H}$'s, to the choice of priors. Therefore, understanding
the theoretical properties of $\mathrm{C}_{3}\mathrm{H}$'s (BI3) and (BI4) could
be an
important step toward establishing a general scheme for dealing with
nuisance parameters in the context of measuring the fraction of
missing information.

%s5 ###
\section{Possible Thesis Topics}
As we concluded in our paper, much remains to be done, especially with
small sample sizes. The three discussions vividly demonstrate this,
and point clearly to a number of concrete research directions. Here
are a few possible thesis titles inspired by the discussions:

\begin{enumerate}
\item[$\bullet$]\textit{On Optimal Follow-up Designs in Genetic Hypothesis Testing
Problems}.
\item[$\bullet$]\textit{Measuring Uncertainty in Relative Information Estimation}.
\item[$\bullet$]\textit{On Measuring Relative Information for Semiparametric
Models.}
\item[$\bullet$]\textit{Measures of Information for Artificial Likelihoods.}
\item[$\bullet$]\textit{Implementing Bayesian Relative Information Measures for
Designing Infectious Disease Studies.}
\item[$\bullet$]\textit{Optimal Design Strategies for Testing Regression
Functions Under Constraints.}
\item[$\bullet$]\textit{Dealing with Nuisance Parameters in Measuring the Fraction of
Missing Information.}
\end{enumerate}

Some of these topics are middle-hanging fruits waiting to be picked, so
if you are a
thesis-topic seeking student reading this set of discussions in the
reverse order, go to the first page as soon as possible!

\section*{Acknowledgments}
We thank Yves Chretien for proofreading and suggestions, David
Harrington for discussions on Cox regression, Peter McCullagh for
very helpful\break exchanges that led us to discover an incorrect
explanation in an early version of our rejoinder, and Michael Stein
for exchanges on the design issues underlying $\mathrm{C}_{3}\mathrm{H}$'s setting.
\iffalse
\fi

%
\end{document}